**Optical gating and streaking of free-electrons with attosecond precision**

M. Kozák[1,*], J. McNeur[1], K. J. Leedle[2], N. Schönenberger[1], A. Ruehl[3], I. Hartl[3], J. S. Harris[2,4], R. L. Byer[4] and P. Hommelhoff[1,5]

[1] Department of Physics, Friedrich-Alexander-Universität Erlangen-Nürnberg (FAU), Staudtstrasse 1, 91058 Erlangen, Germany, EU

[2] Department of Electrical Engineering, Stanford University, Stanford, California 94305, USA

[3] Deutsches Elektronen-Synchrotron DESY, D-22607 Hamburg, Germany, EU

[4] Department of Applied Physics, Stanford University, Stanford, California 94305, USA

[5] Max-Planck-Institute for the Science of Light, Günther-Scharowsky-Str. 1, 91058 Erlangen, Germany, EU

[*] e-mail: martin.kozak@fau.de

**Contemporary physics and chemistry aim to record phenomena occurring on the time and length scales of electronic dynamics in atomic systems [1-4]. Recent progress in the generation of attosecond XUV pulses [5-7] allows for direct insight into ultrafast processes in the sub-femtosecond regime. Pulsed electron beams further enable atomic spatial resolution via ultrafast electron diffraction and microscopy [8], but with temporal resolution so far limited to hundreds of femtoseconds [9,10]. In this paper we present proof of principle experiments of an optical gating concept for free electrons. We demonstrate a temporal resolution of 1.2±0.3 fs via energy and transverse momentum modulation as a function of time. The scheme is based on the synchronous interaction between electrons and the near-field mode of a dielectric nano-grating excited by a femtosecond laser pulse with an optical period duration of 6.5 fs. The sub-optical cycle resolution demonstrated here is promising for use in laser-driven streak cameras for attosecond temporal characterization of bunched particle beams as well as time-resolved experiments with free-electron beams. We expect that 10 as temporal resolution will be achieved in the near future using such a scheme.**

The direct observation of electron dynamics in atomic systems has been a long-standing dream of many scientists, leading to the development of advanced experimental techniques such as attosecond optical spectroscopy [5,11,12], high-harmonic orbital tomography [1,13], laser-induced electron self-diffraction [14], and ultrafast electron diffraction [2,8,9]. The latter enables atomic spatial resolution by employing free electron bunches at energies of 30-200 keV with de-Broglie wavelengths on the order of several picometers. Its temporal resolution can be pushed towards the atomic unit of time (24 as) by the experimental technique presented here.

A natural way to manipulate free electrons on the attosecond time scale would be by using the electromagnetic field of a laser pulse oscillating with a period of several femtoseconds. Similarly to optical streaking experiments [10,15], attosecond temporal resolution may be obtained by mapping time to electron energy. When nonlinear forces can be neglected, however, no net energy gain can be imparted to a charged particle by a light wave in vacuum because of energy and momentum conservation. For efficient energy exchange, the symmetry of the field must be broken via the introduction of a boundary surface [10,16].

The laser-driven gating and streaking of electrons presented in this paper is based on the energy and momentum exchange between a laser field and an electron in the proximity of a periodic nanostructure utilizing the inverse Smith-Purcell effect [17-22]. A silicon nano-grating serves as a phase mask for incoming laser pulses, leading to the excitation of a travelling evanescent mode propagating parallel to the electron beam with a phase velocity equal to the electron velocity $v_e=\beta c$ [20-23], where $c$ is speed of light. The direction and strength of the resulting electromagnetic force depends on the optical phase of the excited mode with respect to the arrival time of electrons [23] (see Figure 1a). Spectral filtering of the electrons after interaction with the laser fields yields temporal gating of the electron bunches with sub-optical cycle temporal resolution.

The experimental setup for demonstration of laser-driven gating is shown in Figures 1b, c (for details see section Materials and Methods). A sub-relativistic DC electron beam with an initial energy of $E_{k0}$=28.1 keV ($\beta$=0.32) is focused close to the surface of the grating, parallel to the grating vector. Two infrared femtosecond laser pulses (central wavelength of $\lambda$=1.93 μm, pulse length of $\tau$=600 ± 50 fs and maximum incident field of $E_0$=1.3 ± 0.1 GV/m) with an adjustable relative time delay are focused onto the grating surface at different locations along the electron beam path. After interaction with the grating near fields, the electrons reach an energy high-pass filter transmitting only those with energies $E_k>eU_s$, where $e$ is the electron charge and $U_s$ is the applied retarding voltage.

Depending on their arrival time at the grating with respect to the phase of the accelerating field, electrons are either accelerated, decelerated or deflected (see the electric field and the final momentum change of the travelling electron in Figure 1a) [23]. In our experiment we first imprint a periodic energy modulation to the electron beam via interaction with the first laser pulse. By adjusting the time delay $\Delta t$ of the second laser pulse, this electron energy modulation can be enhanced for $\Delta t=nT$ or suppressed for $\Delta t=(n+1/2)T$, where $T=\lambda/c$ is the period of the driving laser field and $n \in \mathbb{N}$. When only accelerated electrons are detected, this leads to a current that oscillates as a function of $\Delta t$. The sub-cycle energy modulation of the electron beam is maintained due to the short propagation distance $l=18 \pm 1$ µm between the laser spots (corresponding to a travel time of $190 \pm 10$ fs) and the small relative velocity change of the electrons $\Delta v_e/v_{e0}$<1.2%.

The resulting phase-dependent oscillations of the accelerated electron current are shown in Figure 2a for electrons with minimum energy gain of 1 keV compared to the theoretical model (details are discussed in Simulations and Supplementary Figure 1). The temporal duration of the electron bunch accelerated within one optical cycle of the accelerating mode is apparent from the width of the individual peaks in the phase oscillations. The peak width decreases with increasing spectrometer voltage $U_s$ from $3.2 \pm 0.1$ fs (FWHM) at a minimum energy gain of $(eU_s-E_{k0})$=30 eV to $2.0 \pm 0.3$ fs for $(eU_s-E_{k0})$=1 keV (Figure 2b, detail in Supplementary Figure 3). This is a consequence of the smaller phase-space volume occupied by electrons that gain the highest energy.

As the temporal profile of the electron energy modulation reflects the field of the driving laser pulse, it changes on two different time scales. It reflects both the carrier frequency $\omega$ (phase oscillations) and the envelope function of the laser pulse field $\sim\exp(-2\ln2 t^2/\tau^2)$. To observe the effect of laser pulse envelope on the accelerated population of electrons we monitor the amplitude of the sub-cycle current oscillations (c.f. Figure 2a) as a function of the time delay between the laser pulses (see Figure 2c, d). Here the width of the resulting cross-correlation changes from $490 \pm 30$ fs at $(eU_s-E_{k0})$=30 eV to $180 \pm 40$ fs at $(eU_s-E_{k0})$=1.3 keV. This demonstrates that apart from sub-cycle temporal resolution, the gating scheme can provide an envelope temporal resolution more than 3× finer than the duration of the driving laser pulse.

As can be seen in Figure 2a, this technique generates a pulse train of sub-cycle femtosecond electron bunches. To generate a single bunch, i.e. to demonstrate the feasibility of single-cycle gating operation, we simulate gating of a DC electron beam driven by ultrashort infrared ($\lambda$=2 µm, $\tau$=10 fs corresponding to 1.6 optical cycles) laser pulses with a stable carrier-envelope phase (for details see the supplementary section Simulations).

The calculated accelerated electron current as a function of electron arrival time and minimum energy gain is shown in Figure 3. Single-cycle gating can be achieved by the detection of electrons with energy gains above 100 eV. The physical principle used is analogous to the isolation of single attosecond XUV pulses via spectral filtering of the high-energy part of the high harmonics generated from ultrashort laser pulses [6,24]. The gated electron current depends on the initial electron bunch properties. Assuming an initial bunch length of $\tau_e$=300 fs and a waist radius of the electron beam of $w_e$=70 nm, the single-cycle gated electron current represents 0.3% of the initial electron current, which is highly attractive for time-resolved experiments requiring very high temporal resolution. Note that because of the small transverse beam dimensions we consider 1 electron per pulse or less as an ideal bunch charge. This is in accord with state-of-the-art experiments [25, 26].

Not only sub-cycle temporal gating but also transverse streaking of free electrons is possible by a simple variation of the presented setup (Figure 4a), providing for deflection-based applications and an improvement in temporal resolution. This laser-driven transverse streaking is achieved by using the time-dependent near fields of a nano-grating tilted with an angle of $α=37°$ relative to the axis of electron beam propagation [27]. Electrons thus gain transverse momentum, the magnitude and direction of which depends on the electron arrival time with respect to the optical phase of the laser field and distance from the grating surface. Bunches shorter than 1/4 of the laser optical cycle can be streaked by this scheme with a close-to-linear dependence of the deflection angle on the electron arrival time.

For a demonstration of the dependence of the electron deflection angle on the optical phase of the deflecting fields, we again use two spatially separated laser pulses with an adjustable time delay. The polarizations of both laser pulses are rotated to be parallel to the grating vector. Spatial filtering of the deflected electrons is performed with a metallic knife-edge (see Materials and methods for details).

In Figure 4c we show the measured electron current as a function of the minimum deflection angle and of the phase difference between the two laser pulses. The maximum observed deflection angle of $θ=6.5$ mrad is 6-times higher than the undeflected electron beam divergence angle $θ_0=1.1$ mrad.

Due to the geometry of the tilted grating, the deflecting and accelerating/decelerating forces are coupled. This is advantageous in terms of temporal resolution due to the possibility of simultaneous spatial and spectral filtering of the detected electrons. In Figure 4b we show the phase dependent electron current for electrons with a deflection angle $θ>6$ mrad and an energy gain $\Delta E_k>400$ eV. Double filtering (energy and spatial) further decreases the phase-space region occupied by detected electrons and improves the temporal resolution of sub-cycle gating to $1.3\pm0.3$ fs.

We have thus shown that by utilizing the interaction of electrons with the laser-induced near fields of a nano-grating, we can control the energy and deflected spatial profile of the electron beam with sub-optical cycle precision. These results will lead to breakthroughs in ultrafast electron diffraction and microscopy experiments, an example setup of which we show in Figure 5a. Here an electron bunch first interacts with the accelerating field excited by a few-cycle laser pulse leading to modulation of its kinetic energy $E_k$. The gated electrons are then transmitted and/or diffracted by a sample. Here the dynamics excited by, e.g., a foregoing attosecond XUV pulse (generated via high harmonic generation from a split off fraction of the accelerating laser pulse) can be probed as a function of time delay. Using the detection of electrons with the highest energy gain filtered by an imaging spectrometer, single femtosecond temporal resolution will be achieved. Detection of high energy electrons will also maintain the spatial mode of the electron beam after the gating interaction as they undergo minimal deflection [20,23]. When combined with attosecond XUV pulses generated by the same laser, the optical gating of an electron beam allows for all-optical timing control of the experiment, avoiding the electronic timing jitter usually present in state-of-the-art experiments employing microwave electron bunch compression [28].

Additionally, the angular phase-dependent streaking of electrons demonstrated here can have direct application as a laser-driven analogy of a classical streak camera (Figure 5b) in which the temporal profile of an electron bunch is translated into a transverse spatial distribution. Here, a temporal resolution of 10 as will be directly accessible with available laser and electron beam technology (for details see Materials and methods).

In summary, we have experimentally demonstrated laser-driven temporal gating as well as transverse streaking of free electrons at dielectric nanostructures with single-femtosecond temporal resolution. This technique can be directly applied to ultrafast electron diffraction and microscopy. Moreover, the presented physics is not limited to sub-relativistic electrons [19]. It also enables precise bunch length measurements of relativistic charged particle beams (not only electrons) such as those generated in modern free-electron-laser-based light sources.

## Methods

**Laser and electron beams**

An overview of the optical setup is shown in Supplementary Figure 1. A thulium doped infrared femtosecond fiber laser is used in the experiments. The laser parameters are: pulse length $\tau=600\pm 50$ fs (FWHM), central wavelength $\lambda=1930$ nm, repetition rate of 1 MHz and a maximum pulse energy of 500 nJ. The beam is divided by a polarizing beamsplitter into two arms. The polarization angle of the beam in each arm is adjusted by half-wave plates to be parallel to the grating vector $\vec{K}$ in all experiments. The temporal delay between pulses is controlled by an optical delay stage and two wedge prisms with an opening angle of 0.5° for fine tuning, allowing for a precision of ~200 as. Both laser beams are focused on the silicon grating by a single $f=32$ mm aspherical lens leading to two spots spatially separated about $l=18\pm 1$ μm with the $1/e^2$ radii of $w=7\pm 1$ μm. For the phase-dependent measurements, the temporal delay between laser pulses is first set to 190 ± 10 fs corresponding to the travel time of electrons between the two laser spots and then phase controlled by the wedge prisms. The electron beam is generated by a scanning electron microscope column (Hitachi S-series) and focused to a final $1/e^2$ radius of $w_e$=70 ± 20 nm with a divergence angle $\theta_0$=1.1 ± 0.1 mrad and a probe current $I_p$=3 ± 1 pA. The electron beam energy was adjusted in the range of $E_k$=25-30 keV to match the synchronicity condition in each experiment [23]. The width of the electron beam energy distribution is specified to be ~3 eV.

**Accelerating and streaking element**

Silicon gratings are fabricated from phosphorus-doped silicon substrates with a conductivity of 5-10 ohm-cm. Semiconductor material is used to suppress charging of the grating. However, in the spectral region of our laser operation ($\lambda$=1930 nm), silicon optical properties can be treated as dielectric. Nanopatterning is performed by electron beam lithography (JEOL JBX-6300) and reactive ion etching. The grating is prepared on a mesa structure 50 μm above the substrate surface to minimize electron beam clipping on the substrate edges. In the gating experiment, the grating period is $\lambda_p$=620 nm, the trench depth $t_d$=400 nm and the trench width $t_w$=45% of $\lambda_p$ leading to synchronicity of the fundamental spatial harmonic with $\beta$=0.32 (28 keV) electrons. In the streaking experiment, a grating with a smaller period ($\lambda_p$ =465 nm, $t_d$=400 nm, $t_w$=45% of $\lambda_p$) is used and tilted with respect to the electron propagation direction at an angle of $\alpha$=37° in the *x-z* plane (see Supplementary Figure 1). In this geometry, the fundamental harmonic has a phase velocity synchronous with an electron velocity of $\beta = \lambda_p / (\lambda \cos\alpha) = 0.3$ corresponding to an electron energy of 25 keV.

**Detection setup**

Accelerated electrons are transmitted through a retarding field spectrometer that serves as an energy high-pass filter with applied DC retardation voltage $U_s$. They are then detected by a Chevron-type micro-channel plate detector (MCP) and individual electron counts are temporally correlated with a fast photodiode signal detecting the laser pulses in a time-to-digital converter [29]. The accelerated electron signal forms a peak in the resulting histogram plot of MCP counts vs. temporal delay between each count and the following laser pulse (Supplementary Figure 2). For spatial filtering of the deflected electrons, a metallic knife edge with precise transverse in-situ motorized position control (motion in *x*-direction with precision of 1 μm) is placed in the electron beam 14 ± 0.5 mm downstream of the silicon grating.

**Data processing**

The count rate of accelerated electrons is obtained from each measurement by integrating the counts in the signal peak and subtracting the background count level by the relationship:

$$S = \frac{\int_{t1}^{t2} S(t)dt - \frac{t_2 - t_1}{t_1}\int_0^{t_1} S(t)dt}{t_{int}}, \quad (1)$$

where $S(t)$ is the measured time dependent signal, $t_1$ and $t_2$ represent boundaries of the integration time window which contains the acceleration signal, and $t_{int}$ is the time over which data is collected (see Supplementary Figure 2).

Experimental error of accelerated electron current (error bars shown in Figures 2, Figure 4 and Supplementary Figure 3b) is due primarily to two causes. The statistical error is given by the ratio between the signal count rate $S$ and the statistical noise level $\sqrt{N_{bg}(t_2-t_1)/t_{int}}$, where $N_{bg}$ is the average background count level per bin detected out of the signal time window. The second contribution arises from the pointing instability of the electron beam. The focus position instability of approximately 10 nm causes a signal instability of 10% due to the transverse decay length of acceleration and streaking fields $E(d) \cong \exp(-d/\Gamma)$, where $d$ is transverse distance from the sample surface and $\Gamma=\beta\gamma\lambda/(2\pi)=100$ nm is the decay constant, with $\beta=v_e/c$ and $\gamma = (1-\beta^2)^{-1/2}$.

**Simulations**

Electromagnetic fields of the accelerating mode ($z$-component shown in Figure 1a) are calculated by a finite-difference time-domain (FDTD) technique. In the simulation, the grating has an infinite periodicity in the $z$ direction so that Floquet periodic boundary conditions can be applied. The incident laser pulse is treated as a Gaussian pulse with electric field $E(t) = A\exp(-2\ln 2 t^2/\tau_{int}^2)\exp(i\omega t)$, where $A$ is the field amplitude, $\omega$ is the laser circular frequency and the interaction duration $\tau_{int} = \tau\left[1+(\beta c\tau/w)^2\right]^{-1/2}$ describes the temporal and spatial overlap of the electromagnetic field with electrons moving perpendicularly to the laser propagation with velocity $\beta c$ [23]. The electron beam is described by a classical particle model with Gaussian profile of electron density $n_e = \exp(-2\ln 2 y^2/w_e^2)$ with $w_e$=70 nm and Coulomb repulsion neglected. We note that Coulomb repulsion effects are insignificant in our results due to the low electron current (pA) leading to less than 1 interacting electron per laser pulse. Single electron trajectories are calculated by numerical integration of the relativistic equation of motion with the Lorentz force $\frac{d}{dt}(\gamma m_e \vec{v}_e) = q(\vec{E} + \vec{v}_e \times \vec{B})$ using fields obtained by FDTD calculations, with $\gamma$ the Lorentz factor, $m_e$ the electron mass, $\vec{v}_e$ the electron velocity, $\vec{E}$ the electric field and $\vec{B}$ the magnetic field. To obtain the accelerated electron count rates shown in Figures 2a,b and 3 we numerically integrate the number of electrons with an energy gain $\Delta E_k$ exceeding the difference between the initial electron energy and the spectrometer potential barrier ($eU_s-E_{k0}$). For the calculation of the phase dependent electron count rate in the gating experiment (Figure 2a), two consecutive Gaussian laser pulses with temporal separation $\Delta t$ are used as the incident field. Extensive simulation results are shown in Supplementary Figure 3 compared with measured data.

The theoretical curve shown in Figure 2d, depicting the FWHM of the curves in Figure 2c, is obtained by the following analysis. The accelerating field of both laser pulses is analytically approximated in the electron´s rest frame as:

$$E_a = E_{a0} \exp(-y/\Gamma) \left[ \begin{array}{l} \cos(\omega t_0) \exp\left(-\{v_e(t-t_0)/w\}^2\right) \exp\left(-2\ln 2\{(t-t_0)/\tau\}^2\right) + \\ + \cos(\omega\{t_0-\Delta t\}) \exp\left(-\{[v_e(t-t_0-\Delta t)-\Delta z]/w\}^2\right) \exp\left(-2\ln 2\{(t-t_0-\Delta t)/\tau\}^2\right) \end{array} \right],$$

(1)

where $t_0$ is electron arrival time at the first spot, $\Delta t$ is time delay between laser pulses and $\Delta z$ is spatial separation distance of the two laser spots. The energy gain of individual electrons is obtained by integration of the accelerating field over time in linear approximation. Here the small velocity change of electrons (<1.2 %) during the interaction with the first spot was neglected. Electrons with energy gain above a given threshold are numerically integrated leading to the oscillatory current as function of $\Delta t$.

The results of single-cycle gating of electrons shown in Figure 3 are calculated for a structure containing only one grating tooth (profile shown in Supplementary Figure 4). Assuming the incident laser field is a plane wave and the grating is infinitely periodic in the $z$ direction implies that the accelerating field is infinitely periodic. The energy gain of electrons is then periodic in their arrival time $t_0$ and it is not possible to resolve electrons accelerated by only a single optical cycle of the driving pulse. Hence a grating with at most a few grating teeth is required.

The achievable temporal resolution of our proposed attosecond streak camera is calculated as follows. The transverse angular streaking velocity is defined as $v_s = \dfrac{d\varphi(t_0)}{dt_0}$, where $\varphi(t_0)$ is the deflection angle and $t_0$ the electron arrival time. Angular resolution can be written as $\delta\varphi = \arctan(\delta r/D)$, where $\delta r$=10 μm is spatial resolution of electron detector and $D$=20 cm is the distance between detector and streaking element. The final temporal resolution is then given by $\delta t = \sqrt{\left(\delta\varphi/v_s\right)^2 + \left(\dfrac{dv_s}{dy}\dfrac{w_e}{v_s}\right)^2}$, where $y$ is transverse distance from the grating. For the laser parameters used in the experimental data presented here ($\lambda$=1930 nm, $\tau$=600 fs and a peak field of 1.3 GV/m) and a transmission electron microscope (TEM) with typical electron beam parameters ($E_k$=100 keV, $w_e$=1 nm), the temporal resolution equals $\delta t$=10 as, well below the atomic unit of time (24 as). We note that using double-sided structures, the spatial profile of acceleration mode can be changed to $\cosh(y/\Gamma)$ leading to constant acceleration field in the center of the structure [30]. Such a mode will then even offer 10 as temporal resolution for larger electron beam diameters (10-20 nm).

Acknowledgements

The authors acknowledge founding from ERC grant "Near Field Atto".

Author contributions

M. K. and J. McN. carried out the experiments and designed the silicon gratings. K. J. L. prepared the silicon gratings. N. S., A. R. and I. H. developed the laser pulse compression and provided the laser. M. K. performed the simulations and created the figures. M. K., J. McN. and P. H. interpreted the results and wrote the manuscript. J. S. H., R. L. B. supervised the experiments at Stanford, P. H. in Erlangen. M. K. and J. McN. are co-first authors.

Competing financial interests

The authors declare no competing financial interests.

**Figure 1. Experimental setup used for demonstration of electron gating with sub-cycle resolution. a**, The electric field component $E_z$ (color scale) of the accelerating mode as a function of the electron arrival position within one grating period $\lambda_p$ (arrival time) and transverse distance from the silicon grating (grey structure) surface. Arrows indicate the direction and magnitude of the final momentum change of the travelling electron. **b, c**, Sketch of the experimental setup: a DC electron beam (electrons indicated by spheres) interacts with two subsequent laser pulses (red curves) focused down to radii of $w=7\pm 1$ μm in the vicinity of a Si grating (indicated by gray background pattern). The modulation of electron energy $\Delta E_k=E_k-E_{k0}$ during the interaction with the first laser pulse (visualized by the color scale of the spheres representing the electrons) is further enhanced ($\Delta t=nT$, panel **a**) or canceled ($\Delta t=(n+1/2)T$, panel **b**) by interaction with the second laser pulse depending on the relative phase difference between the laser pulses. Slow electrons are filtered out by an energy high-pass filter so that only accelerated electrons (red) are detected by a micro-channel plate detector (MCP).

**Figure 2. Optical phase-dependent energy modulation of electrons on the 1 fs time scale. a**, Measured phase dependent oscillations of the accelerated electron current for a minimum energy gain ($eU_s$-$E_{k0}$) of 1 keV (squares) compared with the theoretical curve (red), demonstrating phase controlled energy modulation. The theory curve was calculated for the transverse distance of electron beam center from the grating surface $y$=100 nm. The electron current oscillation amplitude was used as a fitting parameter. **b**, FWHM of the individual peaks shown in **a** as a function of ($eU_s$-$E_{k0}$) compared to theory (solid curve). **c**, Phase-dependent oscillation amplitude over a longer time window at ($eU_s$-$E_{k0}$)=30 eV (squares), 1 keV (circles) and 1.3 keV (triangles) fitted by Gaussian functions (curves). The data is vertically shifted for clarity. **d**, FWHM of Gaussian peaks shown in **c** (squares) as a function of ($eU_s$-$E_{k0}$) compared to theory (solid curve).

**Figure 3. Single-cycle optical gating of an electron beam.** Simulated dependence of the accelerated electron current (color scale) on the electron arrival time and the minimum electron energy gain ($eU_s$-$E_{k0}$) for a few-cycle incident laser pulse (see Materials and Methods). Single-cycle gating operation in temporal window of duration $\tau_{s\text{-}c}$=1.2 fs is reached for filtering of electrons with ($eU_s$-$E_{k0}$)>100 eV. The laser pulse parameters used for this simulation are $E_0$=1.3 GV/m, $\tau$=10 fs, $\lambda$=2 μm.

**Figure 4. Optical phase-dependent deflection of electrons on the 1 fs time scale. a**, Sketch of the experimental setup used for demonstration of sub-cycle transverse streaking of electrons. Electron beam (spheres) interacts with the near fields (red regions) of silicon grating (grey) tilted by the angle $\alpha$=37°. Spatially (knife edge) and spectrally (energy high-pass filter) filtered electrons are detected by the MCP. **b**, Measured current of spatially (deflection angle $\theta$>6 mrad, represented by the dashed line in **c**) and spectrally ($\Delta E_k$>400 eV) filtered electrons as a function of time delay $\Delta t$ between laser pulses. Peaks fitted by Gaussian functions (red) with $\tau_{\text{FWHM}}$=1.3 ± 0.3 fs. **c**, Dependence of the measured electron current (color scale) on the minimum deflection angle and the relative phase between the pair of driving laser pulses for electrons with energy gain $\Delta E_k$>30 eV. The full line indicates the electron beam divergence angle.

**Figure 5. Proposed schemes for single-cycle optical gating of electrons for ultrafast electron diffraction and microscopy experiments and a laser-driven attosecond streak camera. a**, Electrons (spheres) coming from the left obtain an energy modulation $\Delta E_k$ by the interaction with an ultrashort laser pulse (red) with optical period $T$ in the vicinity of a single ridge nanostructure (grey). Afterwards they interact with the sample and probe electron dynamics excited by an attosecond XUV pulse (violet). Electrons with the highest energy gain (red) accelerated within one optical cycle are filtered by the spectrometer and detected by a micro-channel plate detector (MCP). With nowadays available parameters, a time resolution below 1 fs is possible. **b**, An electron bunch (spheres) is transversely streaked by the phase-dependent deflecting fields generated by an ultrashort laser pulse (red) and its transverse spatial distribution is detected by a spatially resolved electron detector. Whereas in **a** timing information is mapped onto energy,

here it is mapped to the spatial domain like in a classical streak camera, offering a temporal resolution down to 10 as (see section Simulations).

## Figure 1

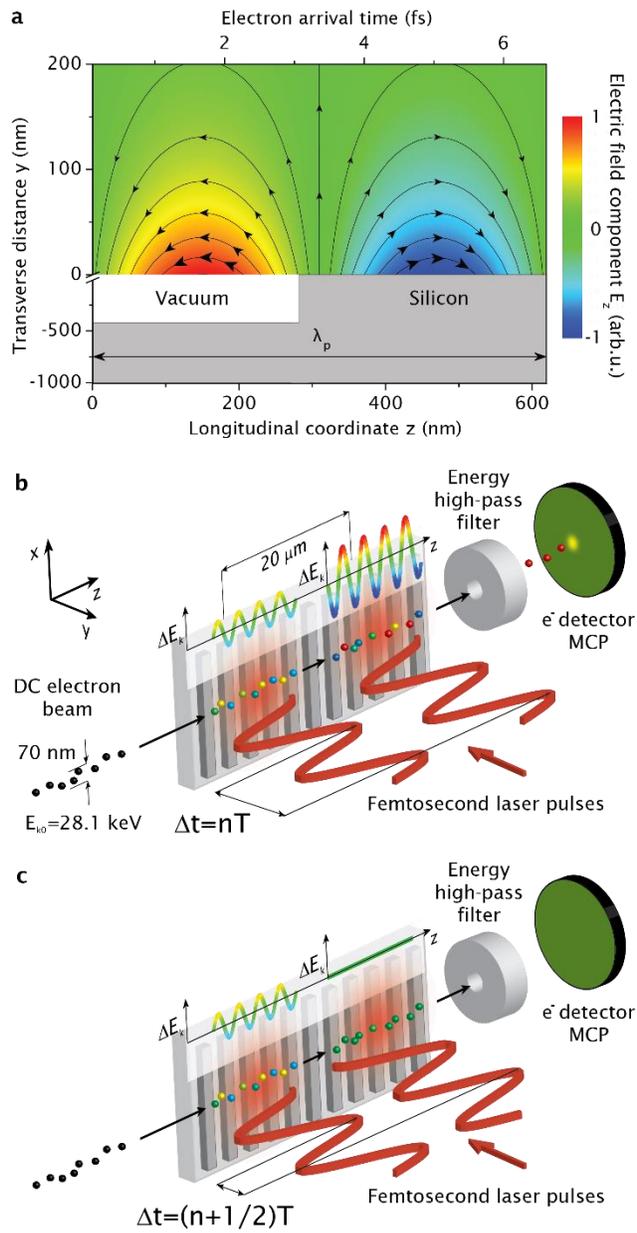

**Figure 2**

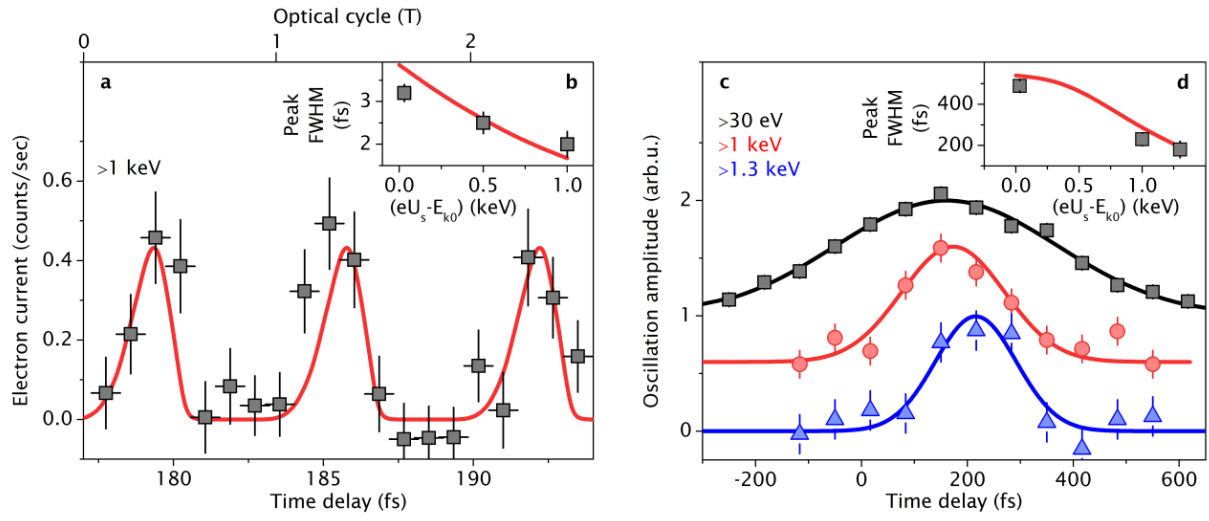

**Figure 3**

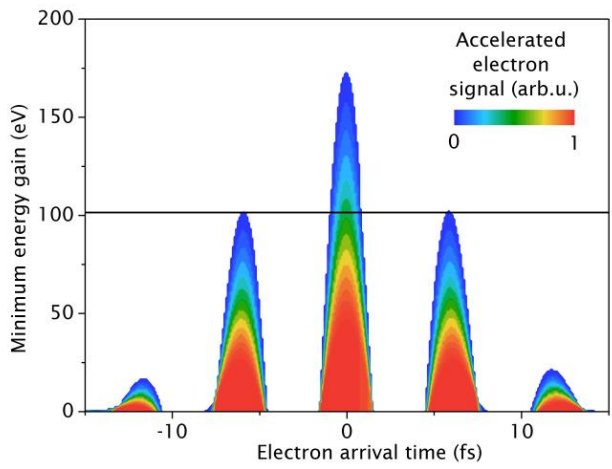

# Figure 4

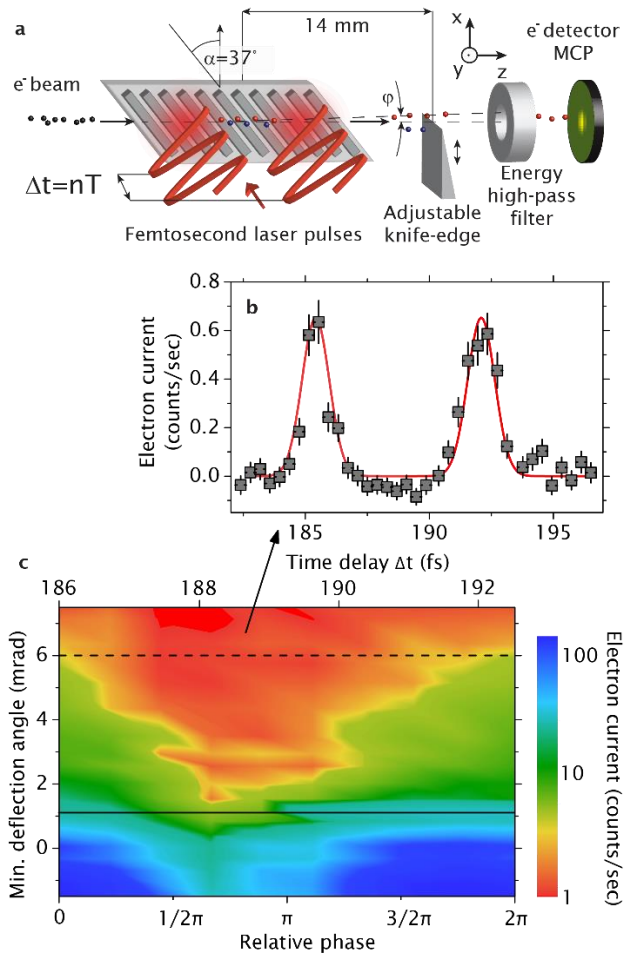

# Figure 5

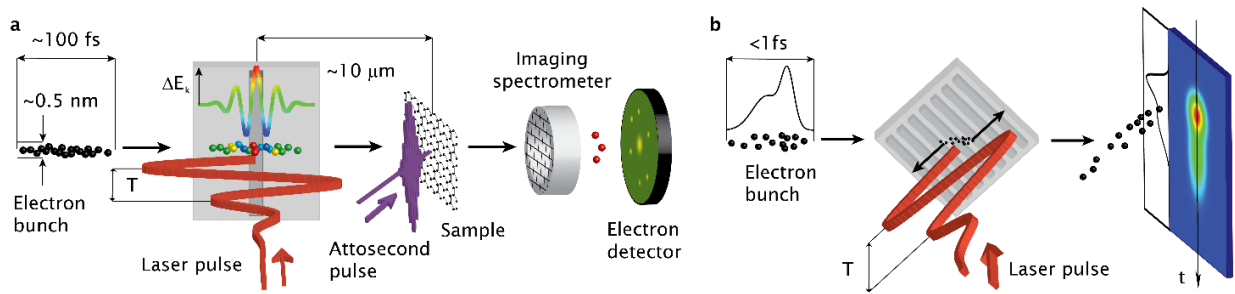